\documentclass[preprint, showpacs, nofootinbib, aps, prd]{revtex4}

\usepackage{graphicx}
\usepackage{dcolumn}
\usepackage{bm}
\usepackage{hyperref}

\begin{document}

\title{An Implication on the Pion Distribution Amplitude from the Pion-Photon Transition Form Factor with the New BABAR Data}

\author{Xing-Gang Wu}
\email[email: ]{wuxg@cqu.edu.cn}
\affiliation{Department of Physics, Chongqing University, Chongqing 400044, P.R. China}
\author{Tao Huang}
\email[email: ]{huangtao@ihep.ac.cn}
\affiliation{Institute of High Energy Physics and Theoretical Physics Enter for Science Facilities, Chinese Academy of Sciences, Beijing 100049, P.R. China}

\date{\today}

\begin{abstract}

The new BABAR data on the pion-photon transition form factor arouses people's new interests on the determination of pion distribution amplitude. To explain the data, we take both the leading valence quark state's and the non-valence quark states' contributions into consideration, where the valence quark part up to next-to-leading order is presented and the non-valence quark part is estimated by a phenomenological model based on its limiting behavior at both $Q^2\to 0$ and $Q^2\to\infty$. Our results show that to be consistent with the new BABAR data at large $Q^2$ region, a broader other than the asymptotic-like pion distribution amplitude should be adopted. The broadness of the pion distribution amplitude is controlled by a parameter $B$. It has been found that the new BABAR data at low and high energy regions can be explained simultaneously by setting $B$ to be around $0.60$, in which the pion distribution amplitude is closed to the Chernyak-Zhitnitsky form. \\

\end{abstract}


\maketitle

\section{Introduction}

The pion-photon transition form factor $\gamma\gamma^*\to \pi^0$, which relates two photons with one lightest meson, is the simplest example for the perturbative application to exclusive processes. It provides a good platform to study the property of pion distribution amplitude (DA), i.e. one can extract useful information on the shape of the leading-twist pion DA by comparing the estimated result on the transition form factor $F_{\pi \gamma}(Q^2)$ with the measured one. The CELLO collaboration has measured the pion-photon transition form factor a long time ago, where, one of the photons is nearly on-shell and the other one is off-shell with a virtuality in the range of low energy region ($Q^2<3$ GeV$^2$) \cite{CELLO}. Later on, the CLEO collaboration also measured such form factor but with a broader range of $Q^2\in$ $[1.5,9.2]$~GeV$^2$ \cite{CLEO}. Very recently, BABAR collaboration does a more precise measurement at both low and high energy region, and their data shows that in the range of $Q^2\in [4,40]$~ GeV$^2$, the pion-photon transition form factor behaves as \cite{babar}
\begin{equation}\label{exp}
Q^2 F_{\pi \gamma}(Q^2) = A\left(\frac{Q^2}{10GeV^2}\right)^{\beta},
\end{equation}
where $A=0.182\pm0.002$ and $\beta=0.25\pm0.02$. Such large $Q^2$ behavior contradicts the well-known asymptotic prediction \cite{lb}, i.e. $Q^2 F_{\pi \gamma} (Q^2)$ tends to be a constant ($2f_\pi$) for asymptotic DA $\phi_{as}(x,Q^2)|_{Q^2\rightarrow \infty}=6x(1-x)$, where the pion decay constant $f_{\pi}=92.4 \pm 0.25$~MeV \cite{pdg}. By extending the previous next-to-leading order (NLO) corrections \cite{asDANNLO0,asDANNLO1} to the present large $Q^2$ region, or even by including the next-to-next-to-leading order corrections \cite{asDANNLO2,asDANNLO3}, the significant growth of the pion-photon transition form factor between $10$ and $40$ GeV$^2$ cannot be explained by using the asymptotic or asymptotic-like DA.

Therefore, many attempts have been tried to solve the present puzzle, some authors have been argued that in contrary to the conventional adopted asymptotic-like DA, the pion DA should be quite broad or even flat in its whole region \cite{broadDA1,broadDA2,broadDA3,broadDA4}. More explicitly, with a flat DA $\phi(x)\equiv 1$, Ref.\cite{broadDA1} shows that the present BABAR data at large $Q^2$ can be explained by choosing proper values for the phenomenological  parameters for the logarithmic model and the Gaussian model constructed there. However, there is no strong reason to support such a flat DA, since the introduced infrared regulator $m^2$ (or $\sigma$) is rightly fitted by the BABAR data. Moreover, one may observe that Ref.\cite{broadDA1} fails to explain the small $Q^2$-behavior, and it can not reproduce the well-known value of $F_{\pi \gamma}(Q^2=0)=1/(4\pi^2 f_\pi)$ that is derived from measuring the rate of $\pi^0\to\gamma\gamma$ \cite{bhl}. Also it can be easily seen that the flat DA with the wavefunction model suggested in Ref.\cite{broadDA1} can not derive the right behavior at $Q^2\to 0$, since as will be shown later it will lead to the probability of finding the valence quark state in the pion, $P_{q\bar{q}}=\int_0^1 \left(\frac{\pi^2 f_{\pi}^2} {3x(1-x)\sigma}\right) dx$, and the charged mean squared radius, $\langle r^2_{\pi^+}
\rangle^{q\bar{q}}=\int_0^1 \left(\frac{\pi^2 f_\pi^2}{2x^2\sigma^2}\right) dx$, both of which are divergent. Furthermore, with such a flat DA, the end-point singularity shall be emerged in many exclusive processes, such as $B\to$ light meson transition form factors, which makes them not calculable in perturbative QCD. This shall greatly compress the applicability of perturbative QCD \footnote{Within the $k_\perp$ factorization approach, by keeping the transverse momentum dependence consistently and with the help of the Sudakov and threshold resummation, this end-point singularity may be cured to a certain degree, e.g. for pion-photon transition form factor \cite{hnli2} and for $B\to$ light form factors \cite{blight}.}.

At present, there is no definite conclusion on whether pion DA is in asymptotic form \cite{lb}, in Chernyak-Zhitnitsky (CZ) form \cite{cz} or even in flat form \cite{flatda}. The pion DA can be expressed in Gegenbauer expansion \cite{lb}. The value of the Gegenbauer moments have been studied in various processes, cf.
Refs.\cite{a2a40,a2a41,a2a42,a2a43,a2a44,a2a45,a2lattice1,a2lattice2,a2lattice3}. The lattice result of Ref.\cite{a2lattice3} prefers a narrower DA with $a_2(1\;{\rm GeV}^2)=0.07(1)$, while the lattice results \cite{a2lattice1,a2lattice2} prefer broader DA, i.e. they obtain $a_2(1\;{\rm GeV}^2)=0.38\pm0.23^{+0.11}_{-0.06}$ and $a_2(1\;{\rm GeV}^2)=0.364\pm0.126$ respectively. These references favor a positive value for $a_2(1\;{\rm GeV}^2)$ and the most recent one is done by Ref.\cite{a2a44}, which shows that $a_2(1\;{\rm GeV}^2)=0.17^{+0.15}_{-0.17}$ through a QCD light-cone sum rule analysis of the semi-leptonic $B\to\pi$ weak transition form factor based on the BABAR data on $B\to\pi l \nu$ \cite{babarbpi}. The pion-photon transition form factor being involved only one pion DA maybe helpful to clarify the present situation.

As argued in Ref.\cite{bhl}, the leading Fock state contributes to $F_{\pi\gamma}(0)$ only half and the remaining half should be come from the higher Fock states as $Q^2\to 0$. And then both contributions from the leading Fock state and the higher Fock states are needed to get the correct $\pi^0\to\gamma\gamma$ rate. In Ref.\cite{huangwu}, we have made such a comprehensive analysis of the pion-photon transition form factor in a smaller $Q^2$ region, e.g. $Q^2\in [0,10]$ GeV$^2$, by taking both the valence quark and the non-valence quark contributions into consideration. It has been found that both the asymptotic-like and the CZ-like DAs can explain the CELLO and CLEO data \cite{CELLO,CLEO} by setting proper parameters for the pion wavefunction. Then it shall be interesting to extend our previous analysis to higher $Q^2$ region so as to determine which pionic behavior is more preferable for consistently explaining the CELLO, CLEO and BABAR data within the whole measured energy region.

The paper is organized as follows. In Sec.II, we present the calculation technology to derive the valence and non-valence contributions to the pion-photon transition form factor. For such purpose, we construct a pion wavefunction model based on the BHL-prescription and present all the necessary formulae for discussion its properties. In Sec.III, we discuss what we can learn of the pionic leading Fock-state wavefunction/DA in comparison with CELLO, CLEO and BABAR experimental data. Some further discussion and comments are made in Sec.IV.

\section{Calculation Technology}

\begin{figure}
\centering
\includegraphics[width=0.70\textwidth]{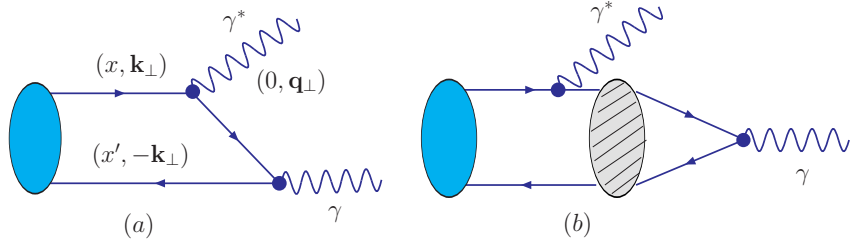}
\caption{Typical diagrams that contribute to the pion-photon transition form factor $F_{\pi \gamma}(Q^2)$, where $x'=(1-x)$. The rightmost shaded oval with a slant pattern stands for the strong interactions.} \label{feyn}
\end{figure}

Generally, the pion-photon transition form factor $\gamma\gamma^*\to \pi^0$ can be written as
\begin{equation}
F_{\pi \gamma}(Q^2)=F^{(V)}_{\pi\gamma}(Q^2)+ F^{(NV)}_{\pi\gamma}(Q^2) ,
\end{equation}
where $F^{(V)}_{\pi\gamma}(Q^2)$ is the usual valence quark part, $F^{(NV)}_{\pi\gamma}(Q^2)$ stands for the non-valence quark part that is related to the higher Fock state of pion. The valence quark contribution $F^{(V)}_{\pi\gamma}(Q^2)$ dominates only as $Q^2$ becomes very large. Fig.(\ref{feyn}) shows this point more clearly. $F^{(V)}_{\pi \gamma}(Q^2)$ comes from Fig.(\ref{feyn}a), which involves the direct annihilation of $(q\bar{q})$-pair into two photons, i.e. the leading Fock-state contribution that dominates the large $Q^2$ contribution. $F^{(NV)}_{\pi \gamma}(Q^2)$ comes from Fig.(\ref{feyn}b), in which one photon coupling `inside' the pion wavefunction, i.e. strong interactions occur between the photon interactions that is related to the higher Fock states' contributions \cite{rady2}. Under the light-cone perturbative QCD approach \cite{lb}, we can obtain the valence part $F^{(V)}_{\pi\gamma}(Q^2)$. While for the non-valence part $F^{(NV)}_{\pi\gamma}(Q^2)$, because of its non-perturbative nature, we shall construct a phenomenological model based on limiting behavior at $Q^2\to 0$ and $Q^2\to\infty$ to estimate it's contribution.

Since the pion wavefunction is the key component of the pion-photon transition form factor, in the following subsections, we shall first make a discussion on its explicit form.

\subsection{Pion Wavefunction and Related DA}

\begin{table}
\caption{The explicit form of the spin-space wavefunction $\chi^{\lambda_{1}\lambda_{2}}(x,{\bf k}_{\perp})$. }
\begin{center}
\begin{tabular}{|c||c|c|c|}
\hline\hline ~~~$\lambda_1\lambda_2$~~~ & ~~~$\downarrow\downarrow (\uparrow\uparrow)$~~~&
~~~$\uparrow\downarrow$~~~& ~~~$\downarrow\uparrow$~~~  \\
\hline $\chi^{\lambda_{1}\lambda_{2}}(x,{\bf k}_{\perp})$ & $-\frac{k_x \pm i k_y} {\sqrt{2(m^{2}_q+{\bf k}_{\perp}^2)}}$ & $\frac{m_q}{\sqrt{2(m^{2}_q +{\bf k}^2_{\perp})}}$ & $-\frac{m_q}{\sqrt{2(m^{2}_q + {\bf k}_{\perp}^2)}}$ \\ \hline\hline
\end{tabular}
\label{tab0}
\end{center}
\end{table}

Taking into account the Melosh rotation \cite{M1974}, the full form of the pion wavefunction can be written as
\begin{equation}\label{wave}
\Psi_{q\bar{q}}(x,{\bf k}_{\perp})=\sum_{\lambda_{1}\lambda_{2}} \chi^{\lambda_{1}\lambda_{2}}(x,{\bf k}_{\perp}) \Psi^{R}_{q\bar{q}}(x,{\bf k}_{\perp}) ,
\end{equation}
where $\lambda_1$ and $\lambda_2$ are helicity states of the two constitute quarks, $\chi^{\lambda_{1}\lambda_{2}}(x,{\bf k}_{\perp})$ stands for the spin-space wavefunction coming from the Wigner-Melosh rotation. $\chi^{\lambda_{1}\lambda_{2}}(x,{\bf k}_{\perp})$ can be found in Refs.\cite{spin1,spin2,spin3,spin4}, whose explicit form is shown in TAB.\ref{tab0}. $\Psi^{R}_{q\bar{q}}(x,{\bf k}_{\perp})$ stands for the spatial wavefunction, and we adopt the factorized model to do our discussion, which is divided into a $x$-dependence part $\varphi_\pi(x)$ and a ${\bf k}_{\perp}$-dependence part. $\varphi_\pi(x)$ may or may not be the distribution amplitude, which depends on the explicit form of the ${\bf k}_{\perp}$-dependence part. Based on BHL prescription \cite{bhl,spin1,spin2,spin3,spin4}, the spatial wavefunction $\Psi^{R}_{q\bar{q}}(x,{\bf k}_{\perp})$ can be written as
\begin{widetext}
\begin{equation}\label{wave1}
\Psi^{R}_{q\bar{q}}(x,{\bf k}_{\perp})=A\varphi_\pi(x) \exp\left[-\frac{{\bf k}_{\perp}^2 +m_q^2} {8{\beta}^2x(1-x)}\right],
\end{equation}
\end{widetext}
where the $x$-dependence part $\varphi_\pi(x)$ can be expanded in Gegenbauer polynomials, and by keeping its first two terms, we obtain
\begin{widetext}
\begin{equation}
\Psi^{R}_{q\bar{q}}(x,{\bf k}_{\perp})=A\left(1+B\times C^{3/2}_2(2x-1)\right) \exp\left[-\frac{{\bf k}_{\perp}^2 +m_q^2}{8{\beta}^2x(1-x)}\right] ,
\end{equation}
\end{widetext}
where the Gegenbauer polynomial $C^{3/2}_2(2x-1)=(3/2)[5(2x-1)^2-1]$. The typical parameter $B$ determines the broadness of the wavefunction. The normalization constant $A$, the harmonic scale $\beta$ and the light constitute quark mass $m_q$ are constrained by several reasonable constraints. The first is the conventional wavefunction normalization condition
\begin{equation}
\label{Aconstrain} \int^1_0 dx \int_{|\mathbf{k}_\perp|^2<\mu_0^2}
\frac{d^{2}{\bf k}_{\perp}}{16\pi^3}\Psi_{q\bar{q}}(x,{\bf
k}_{\perp}) =\frac{f_{\pi}}{2\sqrt{3}},
\end{equation}
where $\mu_0$ stands for some hadronic scale that is of order ${\cal O}(1~{\rm GeV})$. The second is the constraint derived from $\pi^0\rightarrow \gamma\gamma$ decay amplitude \cite{bhl}
\begin{equation}
\label{Bconstrain} \int^1_0 dx \Psi_{q\bar{q}}(x,{\bf k}_{\perp}=0) =\frac{\sqrt{3}}{f_{\pi}}.
\end{equation}
Further more, $m_q$ should be around the conventional adopted value $0.30$ GeV.

The leading Fock-state pion DA at the scale $\mu_0$ takes the following form
\begin{equation}\label{phimodel}
\phi_\pi(x,\mu_0^2)=\frac{2\sqrt{3}}{f_\pi} \int_{|\mathbf{k}_\perp|^2\leq\mu_0^2} \frac{d^2\mathbf{k}_\perp}{16\pi^3} \Psi_{q\bar{q}}(x,\mathbf{k}_\perp).
\end{equation}
Substituting the wavefunction model (\ref{wave}), we obtain
\begin{eqnarray}
\phi_\pi(x,\mu_0^2) &=& \frac{\sqrt{3}A m \beta}{2\sqrt{2}\pi^{3/2}f_\pi} \sqrt{x(1-x)} \left(1+B\times C^{3/2}_2(2x-1)\right) \cdot \nonumber\\
&& \left( \mathrm{Erf} \left[\sqrt{\frac{m^2+\mu_0^2}{8\beta^2 x(1-x)}}\right]- \mathrm{Erf}\left[\sqrt{\frac{m^2}{8\beta^2 x(1-x)}}\right] \right),  \label{ourphi}
\end{eqnarray}
where the error function ${\rm Erf}(x)$ is defined as $\mathrm{Erf}(x)=\frac{2}{\sqrt{\pi}} \int_0^x e^{-t^2}dt$. Such a DA with $B\to 0$ is asymptotic-like, and with the increment of $B$, it shall be broadened to a certain degree, e.g. when $B\sim 0.6$, it will be CZ-like with a $k_{\perp}$-dependence factor which suppresses the end-point singularity.

The pion DA at any scale $Q^2$ can be derived from the initial DA $\phi_\pi(x,\mu_0^2)$ through QCD evolution. The evolution equation up to order ${\cal O}(\alpha_s)$ takes the following form
\cite{brodsky}
\begin{eqnarray}
x x'Q^2\frac{\partial \tilde{\phi}_\pi(x,Q^2)}{\partial Q^2}=C_F\frac{\alpha_s(Q^2)}{4\pi}
\left\{\int_0^1[dy]V(x,y)\tilde{\phi}_\pi(y,\mu_0)-x x'\tilde{\phi}_\pi(x,Q^2)\right\}, \label{eq:evolution}
\end{eqnarray}
where $[dy]=dy dy'\delta(1-y-y')$, $\tilde{\phi}_\pi(x,Q^2)=\phi_\pi(x,Q^2)/(x x')$ with $x'=1-x$, and
\begin{displaymath}
V(x,y)=2C_F\left[x y'\theta(y-x)\left(\delta_{h_1\bar{h_2}}+\frac{\Delta}{y-x}\right)+(1\leftrightarrow2)\right] ,
\end{displaymath}
where $(1\leftrightarrow2)$ means that all the properties of the first constitute quark should be exchanged to that of second one, and vice versa. $\delta_{h_1\bar{h_2}}=1$ when the two constitute quarks' helicities $h_1$ and $h_2$ are opposite and $\Delta\tilde{\phi}_\pi(y,Q^2)= \tilde{\phi}_\pi(y,Q^2)- \tilde{\phi}_\pi(x,Q^2)$. With this evolution equation, we can take the evolution effects in calculating the pion-photon transition form factor.

Moreover, a solution of Eq.(\ref{eq:evolution}) in Gegenbauer expansion has been derived by Ref.\cite{brodsky}, which takes the following form
\begin{equation}
\phi_\pi(x,Q^2)=6x x' \sum_{n=0}^{\infty} a_n(\mu_0^2)\left(\ln\frac{Q^2}{\Lambda_{QCD}^2} \right)^{-\gamma_n}C^{3/2}_n(2x-1), \label{eq:geg}
\end{equation}
where the Gegenbauer polynomials $C^{3/2}_n(2x-1)$ are eigenfunctions of $V(x,y)$ and the corresponding eigenvalues are the ``non-singlet" anomalous dimensions
\begin{displaymath}
\gamma_n = \frac{C_F}{\beta_0}\left(1+4\sum_{k=2}^{n+1} \frac{1}{k}- \frac{2\delta_{h_1\bar{h_2}}} {(n+1)(n+2)}\right),
\end{displaymath}
where $\beta_0=11-2n_f/3$. The non-perturbative coefficients $a_n(\mu_0^2)$ can be determined from the initial condition $\phi_\pi(x,\mu^2_0)$ by using the orthogonality relations for the Gegenbauer polynomials $C^{3/2}_n(2x-1)$, i.e.
\begin{equation}\label{moments}
a_n(\mu^2_0)=\frac{\int_0^1 dx \phi_{\pi}(x,\mu^2_0)C^{3/2}_n(2x-1)} {\int_0^1 dx 6x(1-x)
[C^{3/2}_n(2x-1)]^2} .
\end{equation}

It should be noted that even though the model wavefunction (\ref{wave1}) is constructed by using only the first two Gegenbauer terms in the longitudinal function $\varphi_\pi(x)$, our present DA $\phi_\pi(x,\mu^2_0)$ as shown by Eq.(\ref{ourphi}) can be expanded in a full form of Gegenbauer series, i.e. both the leading and the higher Gegenbauer terms are there, whose corresponding Gegenbauer moments can be calculated with the help of Eq.(\ref{moments}). As will be shown in the following TAB.\ref{tab1}, the second Gegenbauer moment $a_2(\mu^2_0)$ is close but not equal to the parameter $B$. This shows that the DA $\phi_\pi$ is different from $\varphi_\pi$, which is due to the choice of the BHL-transverse momentum dependence and the consideration of all the helicity components' contributions. While by taking a simpler Gaussian-transverse momentum dependence and by taking only the usual helicity component into consideration, e.g. the transverse momentum dependence $\propto\exp\left(-\frac{k^2_\perp}{2\sigma x(1-x)}\right) $ \cite{broadDA1}, it leads to $\phi_\pi(x,\mu^2_0)\equiv\varphi_\pi(x)$, and  $a_2(\mu^2_0)=B$.

\subsection{$F^{(V)}_{\pi\gamma}(Q^2)$ up to NLO}

Under the light-cone perturbative QCD approach \cite{lb}, and by keeping the $k_\bot$-corrections in both the hard-scattering amplitude and the wavefunction, $F_{\pi \gamma}(Q^2)$ has been calculated up to NLO \cite{bhl,huangwu,huang1,rady1,hnli1,hnli2}. It is noted that for high helicity states $(\lambda_{1}+\lambda_{2}=\pm1)$, since their hard parts are proportional to the small current quark mass, we can safely neglect their contributions. As a combination of the LO part \cite{bhl,huangwu,huang1,rady1} and the NLO part \cite{hnli1,hnli2} that keep the $k_\perp$-dependence in the hard kernel, we can obtain the following formula after doing the integration over the azimuth angle,
\begin{eqnarray}
F^{(V)}_{\pi \gamma}(Q^2)&=& \frac{1}{4\sqrt{3}\pi^2}\int_0^1\int_0^{x^2 Q^2}\frac{dx}{x Q^2}\left[1-\frac{C_F \alpha_s(Q^2)}{4\pi}\left(\ln\frac{\mu_f^2}{xQ^2+k_\perp^2} +2\ln{x}+3- \frac{\pi^2}{3} \right)\right] \cdot \nonumber\\
& & \Psi_{q\bar{q}}(x,k_\perp^2) d k^2_\perp , \label{ffv}
\end{eqnarray}
where $[dx]=dxdx'\delta(1-x-x')$, $C_F=4/3$ and $k_\perp=|\mathbf{k}_\perp|$. $\mu_f$ stands for the factorization scale, and for convenience, we take $\mu_f=Q$ \cite{asDANNLO0,asDANNLO1}. Here, without loss of generality, the usual assumption that the pion wavefunction depending on $\mathbf{k}_\perp$ through $k_\perp^2$ only, i.e. $\Psi_{q\bar{q}}(x,\mathbf{k}_\perp) =\Psi_{q\bar{q}}(x,k_\perp^2)$, has been implicitly adopted.

\subsection{$F^{(NV)}_{\pi\gamma}(Q^2)$}

As for $F^{(NV)}_{\pi \gamma}(Q^2)$, due to its non-perturbative nature, it is hard to be calculated in any $Q^2$ region. As stated in Ref.\cite{bhl}, around the region of $Q^2\sim 0$, we can treat the photon `inside' the pion wavefunction (nearly on-shell) as an external field that is approximately constant throughout the pion volume. And then, a fermion in a constant external field is modified only by a phase, i.e. $S_A(x-y)=e^{-ie(y-x)\cdot A}S_F(x-y)$. Consequently, the lowest $q\bar{q}$-wavefunction for the pion is modified only by a phase $e^{-i e y\cdot A}$, where $y$ is the $q\bar{q}$-separation. Transforming such phase into the momentum space and applying it to the wavefunction, we can obtain the two limiting behavior of $F^{(NV)}_{\pi \gamma}(Q^2)$ at $Q^2\to 0$, which can be written as
\begin{equation}
F^{(NV)}_{\pi \gamma}(0) = F^{(V)}_{\pi \gamma}(0) =\frac{1}{8\sqrt{3}\pi^2}\int dx\Psi_{q\bar{q}}(x,\mathbf{0}_\perp) ,
\end{equation}
and
\begin{eqnarray}
\frac{\partial}{\partial Q^2}F^{(NV)}_{\pi \gamma}(Q^2)|_{Q^2\to 0} &=& \frac{1}{8\sqrt{3}\pi^2} \left[\frac{\partial}{\partial Q^2}\int_0^1\int_{0}^{x^2 Q^2}\left(\frac{\Psi_{q\bar{q}}(x,k_\perp^2)}{x^2 Q^2}\right)dx
dk_\perp^2\right]_{Q^2\to 0} \nonumber\\
&=& \frac{-A}{128\sqrt{3}m^2 \pi^2\beta^2} \int_0^1 \left(1+B\times C^{3/2}_2(2x-1)\right) \frac{x }{x'}(m^2+4x x' \beta^2) \cdot \nonumber\\
&& \exp\left[-\frac{m^2}{8{\beta}^2x x'}\right] dx ,
\end{eqnarray}
where $x'=1-x$. The above equation shows explicitly that at $Q^2\to 0$, the leading Fock state contributes to $F_{\pi\gamma}(0)$ only half, i.e. $F^{(V)}_{\pi \gamma}(0) =F_{\pi\gamma}(0)/2$. While by taking both the valence and non-valence contributions into consideration, one can get the correct rate of the process $\pi^0\to\gamma\gamma$.

Next, we construct a phenomenological model for $F^{(NV)}_{\pi\gamma}(Q^2)$ by requiring it satisfy the above listed two limiting behavior at $Q^2=0$ and by assuming that it is power suppressed to $F^{(V)}_{\pi\gamma}(Q^2)$ in the limit $Q^2\to\infty$. For such purpose, we adopt the model constructed in Ref.\cite{huangwu}
\begin{equation}
F^{(NV)}_{\pi \gamma}(Q^2)=\frac{\alpha}{(1+Q^2/\kappa^2)^2} ,
\end{equation}
where $\kappa=\sqrt{-\frac{F_{\pi\gamma}(0)}{\frac{\partial}{\partial Q^2}
F^{(NV)}_{\pi \gamma}(Q^2)|_{Q^2\to 0}}}$ and $\alpha=\frac{1}{2}F_{\pi\gamma}(0)$. It is easy to find that $F^{(NV)}_{\pi \gamma}(Q^2)$ will be suppressed by $1/Q^2$ to $F^{(V)}_{\pi \gamma}(Q^2)$ in the limit $Q^2\to\infty$. Then at large $Q^2$ region, the non-valence Fock state part $F^{(NV)}_{\pi \gamma}(Q^2)$ shall give negligible contribution to the form factor. However it shall give sizable contribution at small $Q^2$ region.

\subsection{Probability $P_{q\bar{q}}$ and Charged Mean Square Radius $\langle r^2_{\pi^+}\rangle^{q\bar{q}}$ }

After deriving the possible ranges for the parameters in the pion wavefunction, we shall meet the question that whether the resultant wavefunction and hence its DA is reasonable or not. In addition to the pion-photon transition form factor, the pion electromagnetic form factor $F_{\pi^+}(Q^2)$ also provides a platform for studying the properties of pion wavefunction \cite{spin3,pie1,pie2,pie3,pie4}.

For such purpose, following the same procedure as described in detail in Ref.\cite{spin3}, we derive a formula for the soft part contribution by taking all the helicity components' contribution to the pion electro-magnetic form factor. The general form for the soft part contribution can be written as \cite{dy}
\begin{equation}\label{soft}
F^s_{\pi^+}(Q^2)=\int\frac{dx d^{2}{\bf k}_{\perp}}{16\pi^3} \sum_{\lambda_1,\lambda_2} \Psi_{q\bar{q}}^*(x,{\bf k_\perp},\lambda_1)\Psi_{q\bar{q}}(x,{\bf k'_\perp},\lambda_2),
\end{equation}
where $Q^2={\bf q}_{\perp}^2$ and ${\bf k'}_{\perp}={\bf k}_{\perp}+(1-x){\bf q}_{\perp}$ for the final state LC wavefunction when taking the Drell-Yan-West assignment. We can derive the probability for finding the lowest valence quark state $P_{q\bar{q}}$ and the charged mean square radius $\langle r^2_{\pi^+} \rangle^{q\bar{q}}$ from the limiting behavior of $F^s_{\pi^+}(Q^2)$ at $Q^2\to 0$.

Substituting the pion model wavefunction (\ref{wave}) and finishing the integration over $\mathbf{k}_{\perp}$ with the help of the Schwinger $\alpha-$representation method, $\frac{1}{A^{\kappa}}=\frac{1} {\Gamma(\kappa)}\int_0^{\infty} \alpha^{\kappa-1}e^{-\alpha A}d\alpha$, Eq.(\ref{soft}) can be simplified as
\begin{eqnarray}
F^s_{\pi^+}(Q^2)&=&\int^1_0 dx\int^\infty_0 d\lambda \frac{A^2}{128\pi^2 ( 1 + \lambda)^3} \exp\left[-\frac{8m_q^2{( 1 +\lambda) }^2 + Q^2x^{'2}( 2 + \lambda( 4 + \lambda))}{32x' x\beta^2 ( 1 + \lambda) }\right]\times\nonumber\\
& & \left[1+B\times C^{3/2}_2(2x-1)\right]^2 \left\{ I_0\left(\frac{-Q^2 x' \lambda^2} {32x\beta ^2(1 + \lambda)}\right)\bigg[32x' x\beta^2(1 +\lambda) -\right.\nonumber\\
& & \left. Q^2 x^{'2} ( 2 + \lambda ( 4 +\lambda )) + 8m_q^2(1 +\lambda)^2 \bigg]- I_1\left(\frac{-Q^2 x' {\lambda }^2}{32x\beta^2(1 + \lambda) }\right)Q^2 x^{'2} \lambda^2 \right \},
\end{eqnarray}
where $x'=1-x$ and $I_n\,\,(n=0,1)$ stands for the modified Bessel function of the first kind. After taking the expansion in the small $Q^2$ limit, we obtain the probability $P_{q\bar{q}}$ for the valence quark state,
\begin{eqnarray}
P_{q\bar{q}}&=& F^s_{\pi^+}(Q^2)|_{Q^2=0}=\int \frac{dx d^2\mathbf{k}_{\perp}}
{16\pi^3}|\Psi_{q\bar{q}}(x,\mathbf{k}_{\perp})|^2 \nonumber\\
&=&\int^1_0 dx\int^\infty_0 d\lambda \frac{A^2} {16\pi^2 ( 1 + \lambda)^2}\left[1+B\times C^{3/2}_2(2x-1)\right]^2 \exp\left(-\frac{m_q^2 ( 1 + \lambda)}{4 x' x\beta^2}\right) \nonumber\\
& & \Bigg[m_q^2 (1 +\lambda)+4x x'\beta^2\Bigg]
\end{eqnarray}
and the charged mean square radius $\langle r^2_{\pi^+}\rangle^{q\bar{q}}$,
\begin{eqnarray}
\langle r_{\pi^+}^2 \rangle^{q\bar{q}} &\approx&-6\left.\frac{\partial F^s_{\pi^+}(Q^2)} {\partial Q^2}\right|_{Q^2=0} \nonumber\\
&=& \int^1_0 dx\int^\infty_0 d\lambda\frac{3A^2( 2 + 4\lambda+ {\lambda }^2 ) x'} {256 {\pi }^2x{\beta }^2{( 1 + \lambda) }^3}\left[1+B\times C^{3/2}_2(2x-1)\right]^2 \times\exp\left(-\frac{m_q^2( 1 + \lambda) } {4x' x{\beta }^2}\right)\nonumber\\
& & \Bigg[8x' x{\beta }^2 +m_q^2( 1 + \lambda)\Bigg] .
\end{eqnarray}
In the above two equations, one may observe that the terms in the big parenthesis that are proportional to $m_q^2$ come from the ordinal helicity components, while the remaining terms in the big parenthesis are from the higher helicity components.

\section{Numerical results}

We adopt the NLO $\alpha_s(Q^2)$ to do the numerical calculation, i.e.
\begin{equation}
\alpha_s(Q^2)=\frac{4\pi}{\beta_0\ln(Q^2/\Lambda^2_{QCD})}\left[1-\frac{2\beta_1}{\beta_0^2} \frac{\ln[\ln(Q^2/\Lambda_{QCD}^2)]}{\ln(Q^2/\Lambda_{QCD}^2)}\right],
\end{equation}
where $\beta_0=11-2n_f/3$ and $\beta_1=51-19n_f/3$. The value of $n_f$ varies with the energy scale and the value of $\Lambda_{QCD}$ is determined by requiring $\alpha_s(m_{Z^0})=0.1184$ \cite{pdg}, i.e. $\Lambda_{QCD}=0.231$ GeV.

\subsection{Properties of Pion Wavefunction and Pion DA}

\begin{table}
\caption{Pion wave function parameters under the condition of $m_q=0.30$ GeV, and its probability $P_{q\bar{q}}$, charged mean radius $\sqrt{\langle r^2_{\pi^+}\rangle^{q\bar{q}}}$ (unit: $fm$) and the Gegenbauer moments $a_{2,4,6}(\mu_0^2)$.}
\begin{center}
\begin{tabular}{|c||c|c||c|c|c|c|c|c|}
\hline\hline ~~~$B$~~~ & ~$A ({\rm GeV}^{-1})$~& ~$\beta ({\rm GeV})$~& ~$P_{q\bar{q}}$~& ~$\sqrt{\langle r^2_{\pi^+}\rangle^{q\bar{q}}}$~ & ~$a_2(\mu_0^2)$~ & ~$a_4(\mu_0^2)$~ & ~$a_6(\mu_0^2)$~  \\
\hline
~$0.00$~ & ~$25.06$~& ~$0.586$~& ~$63.5\%$~& ~$0.341$~ & ~$0.027$~ & ~$-0.027$~ & ~$-0.016$~ \\
\hline
~$0.20$~ & ~$21.71$~& ~$0.641$~& ~$60.0\%$~& ~$0.358$~ & ~$0.250$~ & ~$-0.025$~ & ~$-0.034$~ \\
\hline
~$0.30$~ & ~$20.26$~& ~$0.668$~& ~$62.0\%$~& ~$0.378$~ & ~$0.362$~ & ~$-0.018$~ & ~$-0.041$~ \\
\hline
~$0.40$~ & ~$18.91$~& ~$0.695$~& ~$66.1\%$~& ~$0.401$~ & ~$0.471$~ & ~$-0.008$~ & ~$-0.047$~ \\
\hline
~$0.60$~ & ~$16.62$~& ~$0.745$~& ~$79.9\%$~& ~$0.451$~ & ~$0.679$~ & ~$0.020$~ & ~$-0.054$~ \\
\hline\hline
\end{tabular}
\label{tab1}
\end{center}
\end{table}

\begin{figure}
\centering
\includegraphics[width=0.50\textwidth]{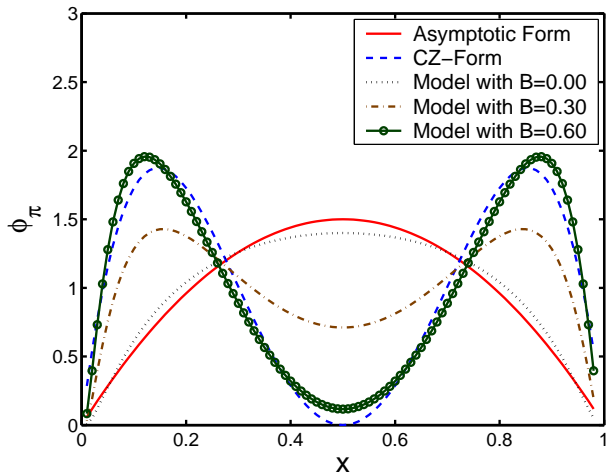}
\caption{Comparison of the pion DA model defined in Eq.(\ref{phimodel}) with the asymptotic-form DA and the CZ-form DA, where $B=0.00$, $0.30$ and $0.60$ respectively. } \label{phi}
\end{figure}

By taking $\mu_0=1$ GeV and $m_q=0.30$ GeV, we present the wavefunction parameters in Tab.\ref{tab1}, which are determined by the mentioned constraints and by taking $B=0.00$, $0.20$, $0.30$, $0.40$ and $0.60$ respectively. The probability for the valence quark state $P_{q\bar{q}}$, the charged mean radius $\sqrt{\langle r^2_{\pi^+}\rangle^{q\bar{q}}}$ (unit: $fm$) and the Gegenbauer moments $a_{2,4,6}(\mu_0^2)$ are also presented in Tab.\ref{tab1}. For the case of $B=0$, $a_{2,4,6}(\mu_0^2)$ can be safely neglected due to their smallness, then the corresponding DA is close to the asymptotic form as shown explicitly by Fig.(\ref{phi}). For a bigger $B$, it is found that $a_{2}(\mu_0^2)$ usually is quite larger than $a_{4,6}(\mu_0^2)$, which is consistent with our model wavefunction (\ref{wave1}), where only the first two Gegenbauer terms are kept in $\varphi(x)$. It is noted that by varying the parameter $B$ within the region of $\sim [0.00,0.60]$, the pion DA shall vary from aymptotic-like to CZ-like form. To show this point more clearly, we draw the pion DA defined in Eq.(\ref{phimodel}) in Fig.(\ref{phi}), where $B=0.00$, $0.30$ and $0.60$  respectively. As a comparison, we also present the conventional asymptotic-form DA, $\phi_{AS}(x)=6x(1-x)$ \cite{lb}, and the CZ-form DA, $\phi_{CZ}(x)=30x(1-x)(2x-1)^2$ \cite{cz}. One may observe from Tab.\ref{tab1} that the value of $\langle r_{\pi^+}^2\rangle^{q\bar{q}}$ increases with the increment of $B$, which runs within the region of $[(0.341 {\rm fm})^2,(0.451 {\rm fm})^2]$ by varying $B\in[0.00,0.60]$. These values are somewhat smaller than the measured pion charged radius $\langle r^2\rangle^{\pi^+}_{expt} =(0.657\pm 0.012\; {\rm fm})^2$~\cite{radius} and $(0.641{\rm fm})^2$~\cite{jlab}, but it is
close to the value as suggested in Refs.\cite{spin3,pole,lattice}. Such smaller $\langle r_{\pi^+}^2\rangle^{q\bar{q}}$ for the leading Fock-state wavefunction is reasonable, since the probability of leading Fock state $P_{q\bar{q}}$ is less than $1$ and is about $60\%-80\%$. This confirms the necessity of taking the higher Fock-states into consideration to give full estimation of the pion electromagnetic form factor/pion-photon transition form factor, especially for lower $Q^2$ regions.

A naive pion wavefunction model has been suggested in Ref.\cite{broadDA1} to explain the new BABAR data \cite{babar}, which is constructed with a flat DA together with a Gaussian ansatz for the $k_\perp$-dependence, i.e.
\begin{equation}
\Psi_{q\bar{q}}(x,\mathbf{k_\perp})=\frac{4\pi^2 f_\pi \phi_\pi(x)}{\sqrt{3}xx'\sigma} \exp\left(-\frac{k^2_\perp}{2\sigma xx'}\right),
\end{equation}
where $x'=1-x$ and $\phi_\pi(x)\equiv1$. With such a model (by setting $\sigma=0.53 {\rm GeV}^2$), it can be easily see that one can not derive the right behavior at $Q^2\to 0$ \cite{broadDA1}, since following the similar steps as shown in Sec.II.D, it will lead to $P_{q\bar{q}}=\int_0^1 \left(\frac{\pi^2 f_{pi}^2}{3xx'\sigma}\right) dx$ and $\langle r^2_{\pi^+}\rangle^{q\bar{q}} =\int_0^1 \left(\frac{\pi^2 f_\pi^2}{2x^2\sigma^2}\right) dx$, both of which are divergent. This in some sense explains why such model wavefunction can explain the pion-photon transition form factor's large $Q^2$ behavior (due to the large enhancement at the end-point region), but fails to explain the lower $Q^2$ behavior.

\begin{figure}
\centering
\includegraphics[width=0.60\textwidth]{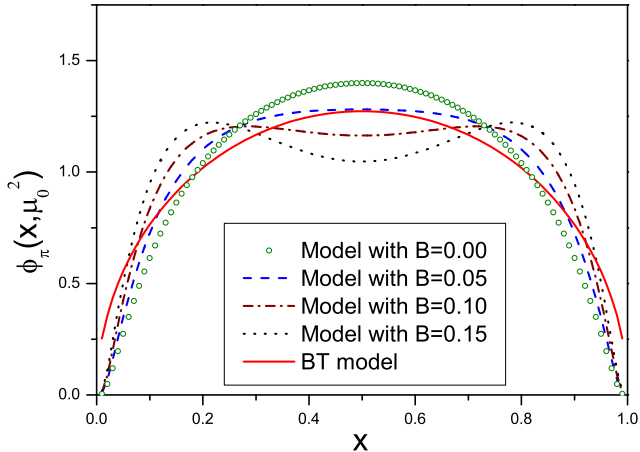}
\caption{Comparison of the pion DA model defined in Eq.(\ref{phimodel}) with the Brodsky and Teramond's holographic model (BT model), where $B=0.00$, $0.05$, $0.10$ and $0.15$ respectively. } \label{phibrod}
\end{figure}

Next, it would be interesting to make a comparison with Brodsky and Teramond's holographic model (BT model) with a quark mass effect \cite{wfbrodsky1,wfbrodsky2} for the pion DA. The BT model is predicted by using the anti-de Sitter / conformal field theory (AdS/CFT) correspondence and by using the soft-wall holographic model, whose explicit form is \cite{wfbrodsky2}
\begin{equation}
\phi_M(x,\mu_0^2)={\cal C}\sqrt{x(1-x)}\exp\left[-\frac{1}{2\kappa^2}\left( \frac{m_u^2}{x} + \frac{m_d^2}{1-x}\right)\right] \left[1-\exp\left(-\frac{\mu_0^2}{2\kappa^2 x(1-x)}\right)\right],
\end{equation}
where $\kappa=0.375$ GeV \cite{wfbrodsky1}, $m_u=2$ MeV and $m_d=5$ MeV \cite{wfbrodsky2}. The factor ${\cal C}\simeq2.55$, which can be determined by its normalization. It is found that for $\mu_0 \sim 1 GeV$, the term involving $\mu_0$ gives quite small contribution and it can be safely neglected as is done by Ref.\cite{wfbrodsky2}. A comparison of our present pion DA model (\ref{phimodel}) with that of the BT model is presented in Fig.(\ref{phibrod}), where our model with $B=0.00$, $0.05$, $0.10$ and $0.15$ are presented by the circles, the dashed, the dash-dot and dotted lines respectively, and the BT model is drawn by a solid line. Since both models have similar transverse momentum behavior, it is natural to estimate that when setting $B\simeq 0.125$, which corresponds to the same second Gegenbauer moment of BT model $a_2(\mu^2_0)\sim 0.145$, these two models shall lead to a similar behavior for the pion-photon transition form factor if calculating under the calculation technology as described in Sec.II.C \footnote{For such a calculation, one needs to be careful that the spin-space wavefunction for BT model should be changed accordingly, since $m_u$ and $m_d$ are taken different values that is different to our present treatment of $m_u=m_d=m_q$. }.

\subsection{Pion-Photon Transition Form Factor}

\begin{figure}
\centering
\includegraphics[width=0.50\textwidth]{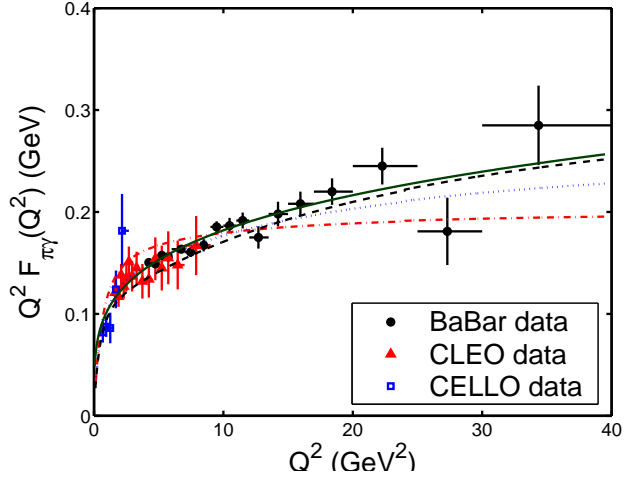}
\caption{$Q^2 F_{\pi\gamma}(Q^2)$ with the model wavefunction (\ref{wave}) by taking $m_q=0.30$ GeV and by varying $B$ within the region of $[0.00,0.60]$. The dash-dot line, the dotted line and the dashed line are for $B=0.00$, $B=0.30$ and $B=0.60$ respectively. The solid line is the fitted curve (\ref{exp}) derived by BABAR \cite{babar}.} \label{figff1}
\end{figure}

First, we calculate the pion-photon transition form factor with the model wavefunction (\ref{wave}) by taking $m_q=0.30$ GeV and by varying $B$ within the region of $[0.00,0.60]$. The result is shown in Fig.(\ref{figff1}), where the dash-dot line, the dotted line and the dashed line are for $B=0.00$, $B=0.30$ and $B=0.60$ respectively. As a comparison, we also present the BABAR fitted curve (\ref{exp}) in Fig.(\ref{figff1}), which is shown by a solid line. For small energy region, $Q^2\lesssim 15~GeV^2$, it is found that both asymptotic-like and CZ-like  wavefunctions by adjusting the quark mass parameter can explain the CELLO, CLEO and BABAR experimental data, which agrees with the observation in Ref.\cite{huangwu}. However, at large $Q^2$ region, different behavior of DA (by varying $B$) shall lead to different limiting behavior. Typically, it is found that when $Q^2\to\infty$, the $Q^2 F_{\pi \gamma}(Q^2)$ for asymptotic-like wavefunction (with $B=0$) tends to the usual limit $2f_\pi\simeq 0.185 GeV$ \cite{lb}. So to explain the newly obtained BABAR data on high energy region, we need a broader DA other than the asymptotic one. It is found that with the increment of $B$ (corresponding to a more broader DA as shown by Fig.(\ref{phi})), the estimated pion-transition form factor shall be more close to the BABAR data. Therefore the pion DA behavior will be determined If BABAR present measurement can be confirmed in the coming future.

\begin{figure}
\centering
\includegraphics[width=0.50\textwidth]{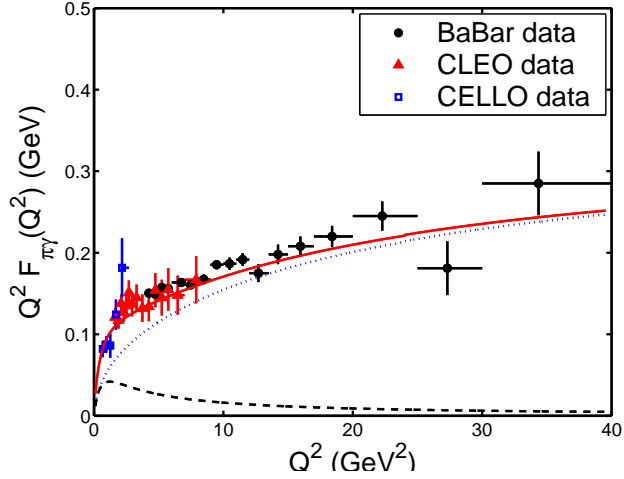}
\caption{$Q^2 F_{\pi\gamma}(Q^2)$ with the model wavefunction (\ref{wave}) by taking $m_q=0.30$ GeV and $B=0.60$. The solid line, the dotted line and the dashed line are for total contribution, the leading valence quark contribution and the  non-valence quark contribution to the form factor respectively.} \label{figff2}
\end{figure}

Second, we show how the leading valence quark and the non-valence quark contribute to the pion-photon transition form factor. We show the results for $B=0.60$ in Fig.(\ref{figff2}), where the solid line, the dotted line and the dashed line are for total contribution, the leading valence quark contribution and the non-valence quark contribution to the pion-photon transition form factor respectively. Fig.(\ref{figff2}) shows that the leading valence Fock-state  contribution dominates the pion-photon transition from factor $Q^2 F_{\pi\gamma}(Q^2)$ for large $Q^2$ region, and the non-valence quark part is small in high $Q^2$ region, but it shall provide sizable contribution to the low and intermediate energy regions. So one should consider the non-valence Fock states' contribution to $Q^2 F_{\pi\gamma}(0)$ so as to explain the experimental data at both low and high $Q^2$ region.

\begin{figure}
\centering
\includegraphics[width=0.50\textwidth]{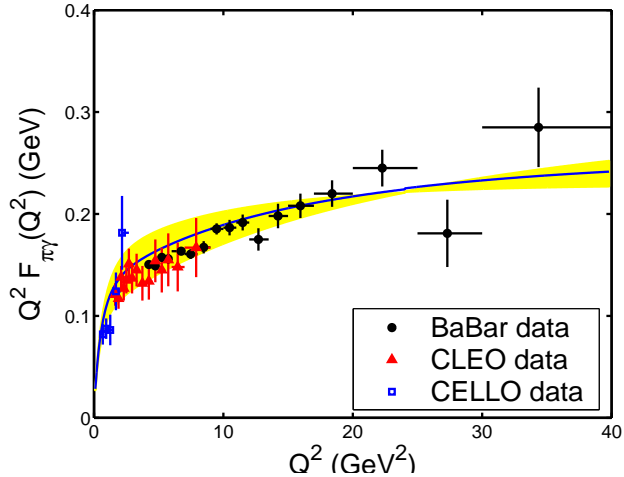}
\caption{$Q^2 F_{\pi\gamma}(Q^2)$ with the model wavefunction (\ref{wave}) by fixing $B=0.60$ and by varying $m_q$ within the region $[0.30,0.50]$ GeV. The solid line is for $m_q=0.40$ GeV, and the shaded band shows its uncertainty.} \label{figff3}
\end{figure}

Third, we make a discussion on the uncertainties caused by varying the value of $m_q$. For such purpose, we fix $B$ to be 0.60 \footnote{The case of $B=0.30$ is similar, only the range of $m_q$ should be shifted to $0.30_{-0.10}^{+0.10}$ GeV \cite{huangwu}. }. From Fig.(\ref{phi}), one may observe that when $B=0.6$, the DA is close to the CZ-form. As has been argued in Ref.\cite{huangwu}, for the case of CZ-like DA, in order to be consistent with the experimental data at low energy scale, $m_q$ should be within the region of $0.40_{-0.10}^{+0.10}$ GeV. So we vary $m_q$ within the region of $[0.30,0.50]$ GeV to show the uncertainties. The results are shown in Fig.(\ref{figff3}), where the solid line is for $m_q=0.40$ GeV, and the shaded band shows its uncertainty. In the lower $Q^2$ region, the upper edge of the band is for $m_q=0.50$ GeV and the lower edge is for $m_q=0.30$ GeV; while in the higher $Q^2$ region, the upper edge of the band is for $m_q=0.30$ GeV and the lower edge is for $m_q=0.50$ GeV.

\section{Summary}

In the present paper, we have taken both the valence quark state's and the non-valence quark states' into consideration. The valence quark part is calculated up to NLO within the $k_T$ factorization approach and the non-valence quark part is estimated by a naive model based on its limiting behavior at both $Q^2\to0$ and $Q^2\to\infty$. Our results show that (1) For  $Q^2\lesssim 15~GeV^2$, it is found that both asymptotic-like and more broader wavefunctions can explain the CELLO, CLEO and BABAR experimental data under reasonable choices of parameters. To be consistent with the new BABAR data at large $Q^2$ region, we need a broader DA, i.e. the conventional adopted asymptotic DA should be broadened to a certain degree. (2) With suitable parameters for the pion model wavefunction that is constructed based on the BHL prescription, it is found that a more broader DA (with larger $B$) shall lead to a better agreement with the BABAR data. If BABAR confirms its present measurement, then pion DA should be broader, such as a CZ-like one with an improved behavior at the end point region. (3) The present adopted model of the pion wavefunction as shown by Eq.(\ref{wave1}) shall present a basis for the application of the pQCD approach \cite{lb,pie1,pie2,pie3,pie4}.

\hspace{1cm}

{\bf Acknowledgments}: The authors would like to thank Dr. Fen Zuo and Dr. Ming-Zhen Zhou for helpful discussions. This work was supported in part by Natural Science Foundation of China under Grant No.10975144, No.10735080 and No.10805082, and by Natural Science Foundation Project of CQ CSTC under Grant No.2008BB0298, and by the Fundamental Research Funds for the Central Universities under Grant No.CDJZR101000616.

\end{document}